\documentclass[10pt,a4paper]{article}
\usepackage{hyperref}

\usepackage[numbers,sort&compress]{natbib}
\usepackage[svgnames]{xcolor}
\usepackage{pst-optexp}
\usepackage{pstricks}
\usepackage{pst-optexp}
\usepackage{pst-all}
\usepackage{pgf,pgfsys,pgffor}

\usepackage{verbatim}
\usepackage{float}
\usepackage{graphics}
\usepackage{graphicx}
\usepackage{amsmath}
\usepackage{amssymb}
\usepackage{latexsym}
\usepackage{color}

\usepackage{caption}
\usepackage{subcaption}
\usepackage{lipsum}

\providecommand{\openone}{\leavevmode\hbox{\small1\kern-4.3pt\normalsize1}}

\begin{document}

\bigskip \thispagestyle{empty}

\begin{center}

\vspace{1.8cm}
\textbf{\large The comparative study of high efficiency of $Tm^{3+}$-doped fiber laser at $1.72\,\mu m$ for different pump schemes}


\vspace{1.5cm}

{\bf Mohamed Zaki}$^{a}${\footnote {\textcolor{blue}{E-mail:\hspace{0.1cm}\href{mailto:mohamed.zaki@edu.uiz.ac.ma}{mohamed.zaki@edu.uiz.ac.ma}}}}, {\bf Mostafa ABOURICHA}$^{a}${\footnote{\textcolor{blue}{E-mail:\hspace{0.1cm}\href{mailto:m.abouricha@uiz.ac.ma}{m.abouricha@uiz.ac.ma}}}} and {\bf Said AMRANE}$^{b}$

\vspace{0.5cm}

$^{a}${\it LPTHE, Department of Physics, Faculty of Sciences, Ibnou Zohr University, Agadir, Morocco.}\\
$^{b}${\it STIC, Department of Physics, Faculty of Sciences, Chouaib Doukkali University, El Jadida, Morocco}

\vspace{1.60cm} \textbf{Abstract}
\end{center}

\baselineskip=18pt \medskip
In this study, we revealed the impact of the pumping scheme, fiber length, pumping power, and reflectivity of the output fiber Bragg grating on the performance of a $Tm^{3+}$-doped fiber laser (TDFL) operating at a wavelength of $1.72\, \mu m$. Using numerical simulations, we optimized the output power and reduced losses due to reabsorption; as well as amplified spontaneous emission (ASE) at approximately $1820\,nm$. The $Tm^{3+}$-doped fiber was bi-directionally pumped at $1570\, nm$ to enhance the pump absorption. The simulations suggest that a maximum power of $5.96 W$ at $1.72\, \mu m$ and a slope efficiency of $64\%$ are achievable using a $Tm^{3+}$-doped silica fiber with a bi-directional pump of $4\, W$ forward and $6\, W$ backward.

\vspace{0.25cm}

\vspace{0.25cm}
{\section{Introduction}}

Fiber laser sources emitting at wavelengths around $1700 nm$, characterized by their high power, have attracted increasing interest due to their potential applications, notably in optical coherence tomography {\color{blue}\cite{sharma2008long,yamanaka2016optical,chong2015noninvasive}}, laser surgery {\color{blue}\cite{wu2015specific}}, remote sensing {\color{blue}\cite{schowengerdt2006remote,emami20171700}}, as well as for methane detection and in the processing of polymeric materials {\color{blue}\cite{mingareev2012welding,anselmo2016gas,chambers2004theoretical}}.

Many laser sources operating at the $1.72\,\mu m$ wavelength have recently been developed; utilizing nonlinear optical frequency conversion methods, including optical parametric oscillators and stimulated Raman scattering{\color{blue}\cite{dong2018high,ma2022robust,zhang2019tunable,pei2021pulsed}}. However, significant advances have been made through the use of rare-earth-doped fibers, which offer a simpler and more robust solution for laser generation at this wavelength. For example, $Tm^{3+}$-doped fiber lasers (TDFL) operating at $1.72\, \mu m$, which utilize energy level transitions within doped fibers, offer a wide emission range from $1650\, nm$ to $2100\, nm$. These lasers can be efficiently pumped by erbium (Er) doped fiber lasers at $1570\,nm$ or Er/Yb co-doped hybrid fibers {\color{blue}\cite{zhang20211}}. Although the first TDFL at $1.72\, \mu m$ was introduced back in 2004 {\color{blue}\cite{agger2004single}}, these high-performance sources represent a recent innovation. This progress has paved the way for the direct use of these high-power fiber lasers, operating at $1.72\, \mu m$, as pumping sources for dysprosium (Dy) doped fibers, enabling the generation of lasers in the mid-infrared range of $3$ to $5\, \mu m$ {\color{blue}\cite{majewski2018emission, majewski2020dysprosium, tang2012study}}.

In recent years, much laser research has focused on improving the output power and efficiency of $Tm^{3+}$-doped fiber lasers at $1.7\, \mu m$. These improvements include the use of fundamental transverse mode fiber laser pump sources from $1.55$ to $1.6\, \mu m$, as well as careful design of the cavity layout {\color{blue}\cite{burns201947, daniel2015ultra}}. An interesting approach was proposed by Zhang et al., who developed an intra-cavity pumping scheme by placing a $1.7\, \mu m$ $Tm^{3+}$-doped fiber cavity inside a $1560\,nm$ Er/Yb-doped fiber cavity. This configuration aims to improve pump absorption and minimize reabsorption loss {\color{blue}\cite{zhang20211}}. In addition, Zhang et al. used a $1720\,nm$ $Tm^{3+}$-doped silica fiber cavity, pumped from both ends by two  Er/Yb-doped cavity fiber lasers {\color{blue}\cite{zhang2022high}}.

In this study, we conducted a comprehensive investigation of the output characteristics of a continuous-wave (CW) thulium-doped fiber laser emitting at $1720\, nm$, which is powered by bidirectional pumping using two $1570\, nm$ sources. The experimental setup, as demonstrated by Zhang et al.{\color{blue}\cite{zhang2022high}}, was described; and a theoretical model based on the rate equations for the thulium-doped fiber was developed. The study then focused on the impact of the distribution of the pump power between the two ends of the fiber and the influence of the reflectivity of the output fiber Bragg grating (FBG$_2$) on the output power, the laser threshold, and the optimal length of the active fiber. The analysis of the results is carried out through numerical simulations.
{\section{Theoretical model}}

{\subsection{Fiber Laser Schematic}}

\begin{figure}[htbp]
    \centering
    \includegraphics[width=0.9\linewidth]{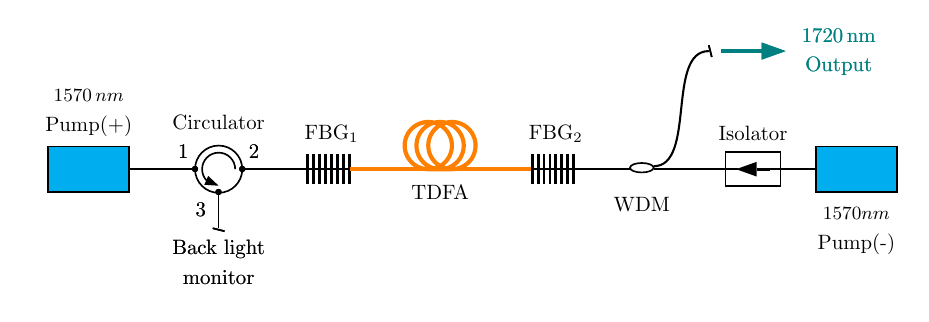}
  \caption{The schematic of the $1720\,nm$ Tm-doped fiber laser.}
  \label{fig:Fig1}
\end{figure}

A schematic diagram of the laser is shown in {\color{blue}Fig.\ref{fig:Fig1}}. The device consists of two $1570\,nm$ Er/Yb-doped fiber lasers, designated as pump sources (pump(+) forward and pump(-) backward),to enhance the injected pumping power. Additionally, a segment of commercial fiber (Nufern SM-TSF-$9/125$) was used as active fiber (TDF) with a low doping concentration $N_{Tm^{3+}}=1.37\times10^{25}\,m^{-3}$ is surrounded by two Fiber Bragg gratings, the first FBG$_{1}$, located at the entrance to the laser cavity, has a reflectivity $R_{s1} >99.5\%$ at $1720\,nm$ and a $3\,dB$ bandwidth of $0.8\,nm$, the second Fiber Bragg grating (FBG$_{2}$), positioned at the cavity exit with a reflectivity $R_{s2}=45\%$ at $1720\,nm$ and a $3\,dB$ bandwidth of $0.2\,nm$. The pump(-) source directs a $1570\,nm$ laser into the cavity using an isolator and a $1570/1720\,nm$ wavelength multiplexer (WDM), which serves to separate the residual $1570\,nm$ pump from the $1720\,nm$ emitted laser. An optical circulator, functioning as an isolator, was installed between pump(+) and FBG$_{1}$ to prevent the return of the residual pump laser. In addition, the free port (port 3) of the circulator is designed for managing the reverse ASE.

{\subsection{Rate equations}}
To find the numerical solutions of a $Tm^{3+}$-doped silica fiber amplifier subjected to pumping in a broadband transition $^{3}H_{6}\xrightarrow{}^{3}F_{4}$ that we have used here (see, for example {\color{blue}\cite{khamis2017enhancement,romano2018simulation, jackson2012towards}}), a model and method are verified in different fiber laser regimes ($Tm^{3+}$-doped fiber {\color{blue}\cite{zhang2022high}}, and Yb-doped fiber {\color{blue}\cite{shang2011comparative, ren2018numerical}}) and are based on the traditional rate equations that have been described in detail in this work ({\color{blue}\cite{peterka2004theoretical,jackson1999theoretical}}) and solved in the steady state considering the CW continuous-wave laser (i.e. time derivatives $\frac{dN}{dt}=0$). The pump power, signal and amplified spontaneous emission (ASE) propagation equations along the active fiber, which can be articulated as follows: {\color{blue}\cite{zhang2022high, zhang2020efficient}}

\begin{equation}
\hspace{-4.8cm}
\frac{\mathrm{d} P_p^{ \pm}(z)}{\mathrm{d} z}= \pm P_p^{ \pm}(z)\left\{\gamma_p\left[\sigma_e\left(\lambda_p\right) N_2(z)-\sigma_a\left(\lambda_p\right) N_1(z)\right]-\alpha_p\right\}
\end{equation}

\begin{equation}
 \hspace{-4.8cm}
\frac{\mathrm{d} P_s^{ \pm}(z)}{\mathrm{d} z}= \pm P_s^{ \pm}(z)\left\{\gamma_s\left[\sigma_e\left(\lambda_s\right) N_2(z)-\sigma_a\left(\lambda_s\right) N_1(z)\right]-\alpha_s\right\}   
\end{equation}

\begin{equation}
\hspace{-0.7cm}
\frac{\mathrm{d} P_{\text{ASE},i}^{\pm}(z)}{\mathrm{d} z} = \pm P_{\text{ASE},i}^{\pm}(z) \left\{\gamma_s\left[\sigma_e(\lambda_i) N_2(z) - \sigma_a(\lambda_i) N_1(z)\right] - \alpha_s\right\} + 2 \sigma_e(\lambda_i) N_2(z) \frac{hc^2}{\lambda_i^3} \Delta\lambda
\end{equation}

Where $P_{p, s, ASE_i}^{+}$ and $P_{p, s, ASE_i}^{-}$ represent the pump power, that of the signal laser and ASE signal at wavelength $\lambda_i$ propagating along the z-axis forward and backward, respectively. In this model, the broadband ASE spectrum $(1650-2100\,nm)$ has been divided into $N_c\approx 400$ channels with a wavelength interval $\Delta\lambda$ of $1.12\,nm$. $\lambda_i$ is the wavelength of the $i^{th}$ channel, and $ASE_i$ is the ASE power at wavelength $\lambda_i$. $\gamma_p$ is the pump overlap factor estimated as the ratio of the core area to the cladding area and $\gamma_s$ the laser signal overlap factor in the $Tm^{3+}$-doped fiber, respectively ; $\alpha_p$ and $\alpha_s$ are the propagation loss coefficients at the pump wavelength and signal wavelength for the $Tm^{3+}$-doped fiber (including background loss and scattering loss), respectively; $\sigma_e(\lambda_i)$ and $\sigma_a(\lambda_i)$ are the $Tm^{3+}$ emission and absorption cross sections taken {\color{blue}\cite{peterka2004theoretical,peterka2011theoretical}}, respectively and in particular $\sigma_a(\lambda_p)$, $\sigma_e(\lambda_p)$, $\sigma_a(\lambda_s)$ and $\sigma_e(\lambda_s)$ are the pump and signal absorption and emission cross sections, respectively; $N_1(z)$ and $N_2(z)$ describe the population densities of ions in the ground state $^3H_6$ and the upper state $^3F_4$ in the longitudinal position $z$ of the fiber, respectively. $N_1(z)$ and $N_2(z)$ are linked by the following two relationships: {\color{blue}\cite{ren2018numerical,cheng2021numerical}}.

\begin{equation}
\frac{N_2(z)}{N_{Tm^{3+}}}=\frac{\frac{\gamma_p}{h c A_{c}} \left[P_p^{+}(z)+P_p^{-}(z)\right]\cdot\lambda_p\sigma_{ap} +\frac{\gamma_s}{h c A_{c}} \sum\limits_{i=1}^{N_c}\left[P^{+}(z, \lambda_i)+P^{-}(z, \lambda_i)\right] \lambda_i \sigma_a(\lambda_i)}{\frac{\gamma_p}{h c A_{c}}\left[P_p^{+}(z)+P_p^{-}(z)\right]\cdot\lambda_p\left(\sigma_{a p}+\sigma_{e p}\right)+\frac{\gamma_s}{h c A_{c}} \sum\limits_{i=1}^{N_c} \left[P^{+}(z, \lambda_i)+P^{-}(z, \lambda_i)\right] \lambda_i \left(\sigma_a(\lambda_i)+\sigma_e(\lambda_i)\right)+\frac{1}{\tau}}  
\end{equation}
\begin{equation}
   N_{Tm^{3+}}= N_1(z) + N_2(z)
\end{equation}

Where $P^{\pm}(z, \lambda_i) = P_{\text{ASE}_i}^{\pm}(z) + P_{s}^{ \pm}(z)$, $N_{Tm^3+}$ is the total population density of $Tm^{3+}$ ions,  $\tau$ is the spontaneous emission lifetime and $A_{c}=\pi.r_c^2$ is the effective doping cross-section area
with $r_c$ is the radius of the dope core.
To simulate the evolution of the signal and pump power and the ASE, we used the fourth-order Runge-Kutta method, introducing the following pump and laser signal boundary conditions:

\begin{equation}
    P_p^{+}(0) =P_{p0}^{+} \hspace{0.5cm};\hspace{0.5cm} P_p^{-}(L)=P_{pL}^{-}
\end{equation}
\begin{equation}
     P_s^{+}(0)=R_{s1} P_s^{-}(0)  \hspace{0.5cm};\hspace{0.5cm} P_s^{-}(L)=R_{s2} P_s^{+}(L)  
\end{equation}
\begin{equation}
     P_{\text {out }}=T_{s2} P_s^{+}(L).\delta
\end{equation}

Where $P_{p0}^{+}$ and $P_{pL}^{-}$ are the pump powers launched into the front end (z = 0) by pump(+) source  and the rear end (z = L) by pump(-) source , respectively; with L is the length of
$Tm^{3+}$-doped fiber; $R_{s1}$ and $R_{s2}$ denote the reflectivity coefficient of the FBG$_1$ and FBG$_2$, respectively; $T_{s2}$ is the FBG$_2$ transmission coefficient such that $T_{s2} = 1-R_{s2}$; $\delta$ indicates the insertion loss of the WDM at $1720 \,nm$ and the output optical isolator. The paraments of the $Tm^{3+}$-doped fiber are summarized in {\color{blue}Tab.\ref{tab:my_label}}.

\vspace{1cm}
\begin{table}[h]
\centering
\begin{tabular}{l l r l}
\hline
\textbf{Quantity, Symbol} & \textbf{Value} & \textbf{Quantity, Symbol} & \textbf{Value} \\
\hline
Pump wavelength, $\lambda_p$ & 1570 nm  & Thulium concentration, $N_{Tm^{3+}}$ & $1.37 \times 10^{25} \, \text{m}^{-3}$ \\
Signal wavelength, $\lambda_s$ & 1720 nm  & Core diameter, $d_c$ & $9 \, \mu \text{m}$ \\
Pump overlap factor, $\gamma_p$ & 0.70  & Clad diameter & 125 $\mu \text{m}$ \\
Signal overlap factor, $\gamma_s$ & 0.70  & Absorption cross section at $\lambda_p$, $\sigma_{ap}$ & $2.50 \times 10^{-25} \, \text{m}^2$ \\
Propagation loss at $\lambda_p$, $\alpha_p$ & $2.50 \times 10^{-2} \, \text{m}^{-1}$ & Emission cross section at $\lambda_p$, $\sigma_{ep}$  & $0.20 \times 10^{-25} \, \text{m}^2$ \\
Propagation loss at $\lambda_s$, $\alpha_s$ & $2.00 \times 10^{-2} \, \text{m}^{-1}$ & Absorption cross section at $\lambda_s$, $\sigma_{as}$ & $3.13 \times 10^{-25} \, \text{m}^2$ \\
Lifetime of the $^3F_4$ level, $\tau$ & 0.25 ms  & Emission cross section at $\lambda_s$, $\sigma_{es}$ & $3.47 \times 10^{-25} \, \text{m}^2$ \\
The insertion loss, $\delta$ & 0.92  & & \\
\hline
\end{tabular}
\caption{Values of numerical parameters. \cite{zhang2020efficient}}
\label{tab:my_label}
\end{table}

\vspace{0.1cm}

{\section{Discussion}}

{\subsection{The effect of pumping scheme}}

\begin{figure}[ht]
  \centering
  \includegraphics[width=0.5\linewidth]{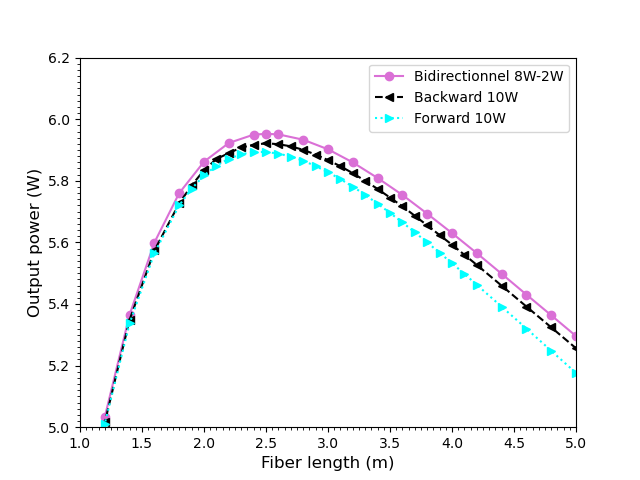}
  \caption{Variation of output power calculated with the $Tm^{3+}$-doped fiber length under different pumping configurations (forward, backward and bidirectional).}
  \label{fig:Fig2a}
\end{figure}

We first study the influence of pumping scheme on amplifier output power, setting the total input pump power at 10 W. The {\color{blue}Fig.\ref{fig:Fig2a}} shows the evolution of output power as a function of fiber length for various pumping schemes (forward, backward, bidirectional). It can be seen that the bidirectional pumping scheme produces significantly higher output power at the optimum fiber length of 2.5 m.

When the active fiber length exceeded the optimum value, the output power at $1720 \,nm$ began to drop rapidly with increasing active fiber length in all three pumping configurations, due to greater cavity loss induced by signal reabsorption.

\begin{figure}[ht]
  \centering
  \begin{minipage}{0.48\linewidth}
    \includegraphics[width=\linewidth]{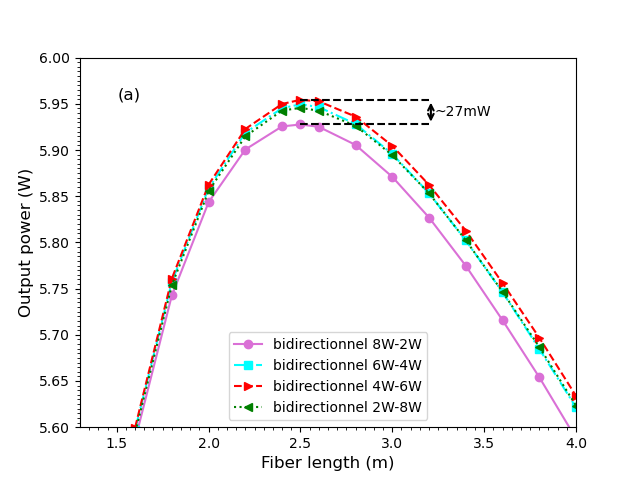}
  \end{minipage}
  \hfill
  \begin{minipage}{0.48\linewidth}
    \includegraphics[width=\linewidth]{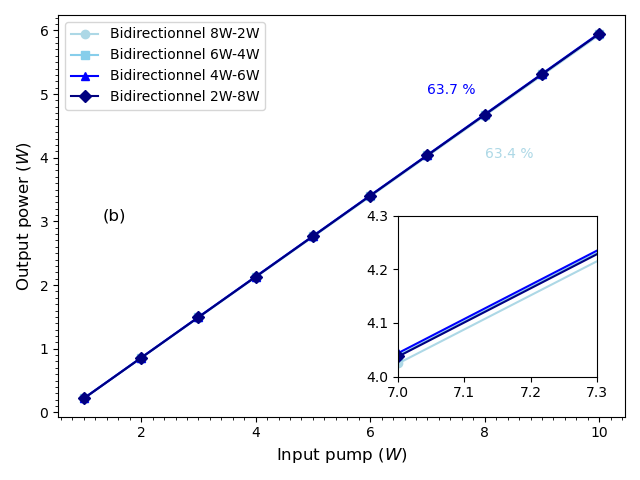}
  \end{minipage}
  \caption{(a) Variation of output power calculated with the $Tm^{3+}$-doped fiber length under different percentages of pump power launched in the case of the bidirectional pumping configuration. b) Output power calculated at 1720 nm as a function of pump power.}
  \label{fig:Fig3}
\end{figure}

Next, we used the bidirectional pumping scheme, varying the percentage of total pumping power in both ends of the fiber to search for optimal values of pump$(+)$ source pumping power $P^+_p(0)$ and pump$(-)$ source pumping power $P^-_p(L)$ while keeping the total power value constant ($P^+_p(0) + P^-_p(L)= 10W)$  as shown in {\color{blue}Fig.\ref{fig:Fig3}a}.

\begin{figure}[ht]
  \centering
  \begin{minipage}{0.48\linewidth}
    \includegraphics[width=\linewidth]{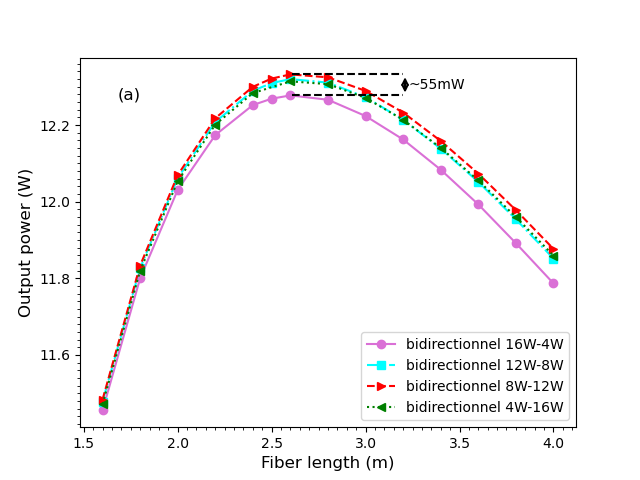}
  \end{minipage}
  \hfill
  \begin{minipage}{0.48\linewidth}
    \includegraphics[width=\linewidth]{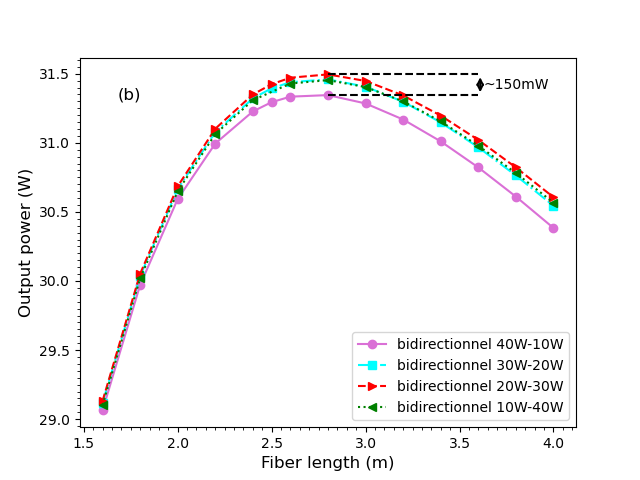}
  \end{minipage}
  \caption{Variation of output power with $Tm^{3+}$-doped fiber length under different pump power percentages in a bidirectional pumping scheme. a) with 20W pump power launched , b) with 50W pump power launched}
 \label{fig:Fig4}
\end{figure}

This analysis shows that when switching from the bidirectional to the bidirectional pumping scheme ($P^{+}_p(0)=4\,W$ from pump$(+)$ source and $P^{-}_p(L)=6\,W$ from pump$(-)$ source), the maximum output power at length $2.5\,m$ shows a minor increase of $\approx27\,mW$ ($+0.46\%$). This increase becomes more significant close to $\approx150\,mW (+0.48\%$) when the pump power is equal to $50\,W$, as shown by \textcolor{blue}{Fig.\ref{fig:Fig4}a} and \textcolor{blue}{Fig.\ref{fig:Fig4}b} where pump powers of $20\,W$ and $50\,W$ are used, respectively.

\textcolor{blue}{Fig.\ref{fig:Fig3}b} illustrates the relationship between the $1720\,\text{nm}$ output power and the $1570\,\text{nm}$ pump power for a $2.5\, \text{m}Tm^{3+}$-doped fiber. A variation in the threshold pump power is observed, shifting from $0.7\,\text{W}$ in an $(8\,\text{W}-2\,\text{W})$ bidirectional pumping configuration with an efficiency of $63.4\%$, to $0.6\,\text{W}$ in a $(4\,\text{W}-6\,\text{W})$ bidirectional pumping configuration, which shows a slightly higher efficiency of $63.7\%$. This means that the $(4\,\text{W}-6\,\text{W})$ bidirectional pumping scheme compares favorably with other bidirectional pumping configurations, delivering considerably higher output power at the optimum fiber length.

\begin{figure}[ht]
  \centering
  \begin{minipage}{0.48\linewidth}
    \includegraphics[width=\linewidth]{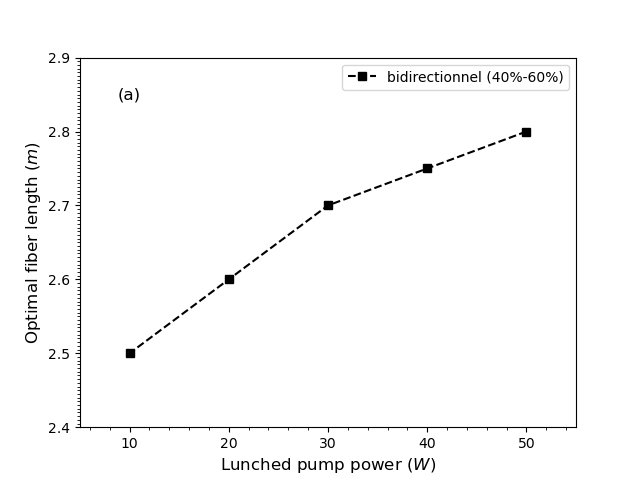}
  \end{minipage}
  \hfill
  \begin{minipage}{0.48\linewidth}
    \includegraphics[width=\linewidth]{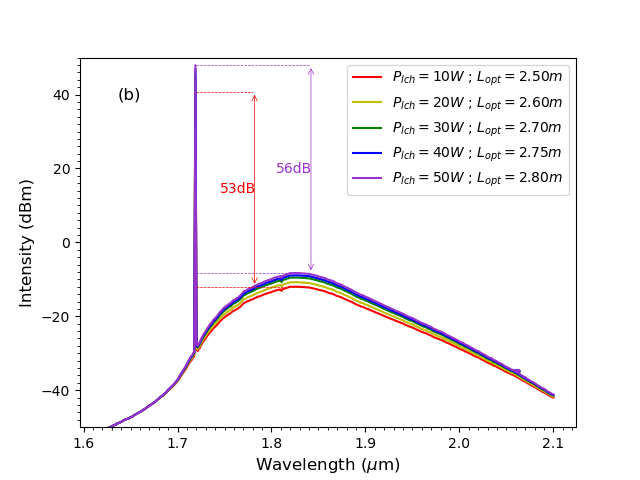}
  \end{minipage}
  \caption{(a) Optimal length of $Tm^{3+}$-doped fiber as a function of applied pump power, and (b) calculated laser spectrum at various pump powers, utilizing the most suitable $Tm^{3+}$-doped fiber length for each selected lanched pump power.}
 \label{fig:Fig5}
\end{figure}

As the total pump power increases, the optimal length of the active fiber is not constant at $2.5m$ and changes in a monotone fashion, as indicated by figures ( {\color{blue}Fig.\ref{fig:Fig3}a}, {\color{blue}Fig.\ref{fig:Fig4}a}, and {\color{blue}Fig.\ref{fig:Fig4}b}, and more specifically {\color{blue}Fig.\ref{fig:Fig5}a}). The increase in optimal length is caused by the low pump reabsorption and the rise in amplified spontaneous emission (ASE) around 1820 nm. When the fiber length exceeds this optimal point, both the output power and efficiency significantly decrease due to the inability to suppress the  $(ASE)$ around $1820nm$. Furthermore, the improvement of the Signal-to-Noise Ratio (SNR) is modest when transitioning from a signal at $1720nm$ with a pump power of $10W (\approx 53dB)$ to a signal with a pump power of $50W (\approx 56dB)$, as indicated by as shown in {\color{blue}Fig.\ref{fig:Fig5}b}.Consequently, selecting a pump power of 10W is deemed suitable to circumvent temperature-related effects. 

{\subsection{The effect of reflectivity}}

\begin{figure}[ht]
  \centering
  \begin{minipage}{0.48\linewidth}
    \includegraphics[width=\linewidth]{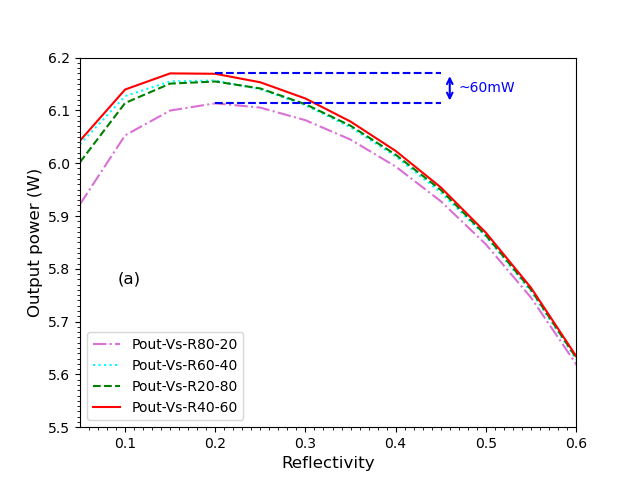}
  \end{minipage}
  \hfill
  \begin{minipage}{0.48\linewidth}
    \includegraphics[width=\linewidth]{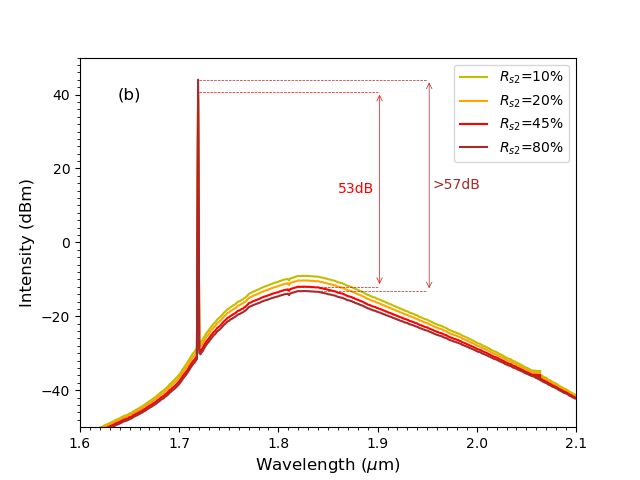}
  \end{minipage}
  \caption{a) Variation of Output Power with the Reflectivity of the Output FBG$_2$ for a $2.5\,m$ $Tm^{3+}$-doped Fiber under Different Pumping Configurations. b) Calculated Laser Spectrum at Various Reflectivities of the Output FBG$_2$, using the Optimal Bidirectional Pumping Configuration $(4\,W-6\,W)$.}
 \label{fig:Fig6}
\end{figure}

This section focuses on studying the influence of the reflectivity of the output Bragg grating (FBG$_2$) on the output performance. According to the results illustrated in {\color{blue}Fig.\ref{fig:Fig6}a}, it is noted that the output power at a wavelength of 1720 nm decreases with an increase in the reflectivity of the FBG. This trend is explained by a limitation in the power extraction rate within the laser cavity. In a bidirectional pumping configuration $(4W-6W)$, the effect of reflectivity on output power is more pronounced, likely due to the increased intensity of the light reflected back towards the source.

On the other hand, using an output FBG with low reflectivity reduces the intra-cavity signal power, which limits the ability to efficiently extract the gain from the $Tm^{3+}$-doped fiber. In this case, the spectral width of the spontaneous emission amplification (ASE) broadens, and its intensity at 1850 nm strengthens, directly influencing the performance of the laser at 1720 nm in terms of the signal-to-noise ratio (SNR) as demonstrated by {\color{blue}Fig.\ref{fig:Fig6}b}. Consequently, to optimize the output power of the laser at 1720 nm, it would be preferable to choose an output FBG with a moderate reflectivity like $R_{s2}=45\%$, which has already been used in previous simulations

\section{Conclusion}

To conclude, this detailed numerical study examines the output characteristics of a thulium-doped silica fiber laser emitting at a wavelength of $1720\,nm$, based on a bidirectional pumping configuration. The analysis focused on the impact of several parameters, including pump power at $1570\,nm$, the length of the $Tm^{3+}$-doped fiber, and the reflectivity of the output fiber Bragg gratings (FBGs). The results reveal a maximum output power of $5.96\,W$, a slope efficiency of $64\%$, and a reduced laser threshold to $0.6\,W$, achieved with a $2.5\,m$ $Tm^{3+}$-doped fiber in a bidirectional pumping setup ($4\,W$ forward and $6\,W$ reverse). We anticipate that these findings will contribute significantly to enhancing the performance of $Tm^{3+}$-doped fiber lasers, particularly around the $1700\,nm$ wavelength.

\addcontentsline{toc}{chapter}{Bibliographie}
\bibliographystyle{unsrt}
\bibstyle{plainnat}
\bibliography{biblio}

\end{document}